\documentclass[twocolumn,showpacs,preprintnumbers,amsmath,amssymb,prl]{revtex4-1}

\usepackage{amssymb}
\usepackage{epsfig,amsmath,graphics}
\usepackage{epstopdf}
\usepackage{graphicx}

\renewcommand{\v}[1]{\ensuremath{\mathbf{#1}}}
\def\be{\begin{equation}}
\def\ee{\end{equation}}

\begin{document}

\title{Atom interferometric gravitational wave detection using heterodyne laser links}

\def\stanfordAffiliation{Department of Physics, Stanford University, Stanford, California 94305}

\author{Jason M. Hogan}
\affiliation{\stanfordAffiliation}
\author{Mark A. Kasevich}
\affiliation{\stanfordAffiliation}

\date{\today}

\begin{abstract}
We propose a scheme based on a heterodyne laser link that allows for long baseline gravitational wave detection using atom interferometry. While the baseline length in previous atom-based proposals is constrained by the need for a reference laser to remain collimated as it propagates between two satellites, here we circumvent this requirement by employing a strong local oscillator laser near each atom ensemble that is phase locked to the reference laser beam.  Longer baselines offer a number of potential advantages, including enhanced sensitivity, simplified atom optics, and reduced atomic source flux requirements.
\end{abstract}

\maketitle

Gravitational wave (GW) detection with atom interferometry offers a promising alternative to traditional optical interferometry \cite{Dimopoulos2008,Hogan2011}. Advantages of this approach include phase multiplication through multiple pulse sequences, proof mass resilience \footnote{The properties of the atomic proof-masses are regular (all atoms are identical) and knowable (well defined magnetic susceptibility).  They are immune from charging events.}, laser frequency noise immunity and quantum back-action noise immunity \cite{Graham2013a}.  Insensitivity to laser frequency noise also allows for the possibility of a detector design based on a single linear baseline, requiring only two satellites instead of the conventional three.

In previous proposals, technical considerations have limited the possible baseline length of the detector.  For suitably low frequency GWs, the sensitivity of the antenna scales linearly with the antenna baseline.  In this Letter, we describe a method which enables antenna operation with substantially longer baselines.  This results in designs whose sensitivities exceed those of existing proposals (e.g. LISA), but which do not require significant advances in the state-of-the-art for atom interferometry.  For example, we describe an antenna with 10 times the sensitivity of the LISA antenna that invokes 12 photon recoil atom optics.

The concept for an atom-based GW antenna is to compare two light-pulse atom interferometers, one at each end of a long baseline.  To implement the atom interferometers, pulses of laser light are used to realize beam splitters and mirrors for the atom de Broglie waves \cite{Kasevich1991}.  In a single-baseline detector, the light pulses are sent back and forth across the baseline from alternating directions, interacting with the atoms on both ends \cite{Graham2013a}.  In this scheme, the phase difference between the atom interferometers is sensitive to variations of the light travel time across the baseline, so by monitoring the phase difference it is possible to detect fluctuations in the light travel time induced by GWs.  Importantly, since the same laser pulses interact with atoms on both sides of the baseline, the common laser phase noise is substantially suppressed in the phase difference between the interferometers \cite{Dimopoulos2008,Graham2013a}.

The envisioned detector consists of two independent spacecraft, each with its own source of ultracold atoms.  The atom interferometers remain inside (or nearby) the local satellite, while telescopes mounted on each satellite are used to send the atom optics laser pulses across the baseline to interact with the atoms on both ends.  Previous GW detection proposals using atom interferometry have assumed that the atom optics laser beam is collimated, constraining the allowed baseline length $L$ to no larger than the Rayleigh range $z_R$ of the laser: $L \leq 2 z_R = 2\pi w^2 / \lambda$, where $\lambda$ is the laser wavelength and $w$ is the radial beam waist \cite{Dimopoulos2008,Graham2013a}.  Assuming $(\Omega/2\pi)\sim \text{kHz}$ Rabi frequencies with $1~\text{m}$ telescopes and $10~\text{W}$ laser power sets a practical limit for baseline length of $L \sim 10^3~\text{km}$ \cite{Dimopoulos2008,Graham2013a}.  By comparison, the traditional LISA design calls for a baseline of at least $10^6~\text{km}$.

Although atom interferometric detectors operating at $1000~\text{km}$ have the potential to reach comparable sensitivity to LISA \cite{Dimopoulos2008}, the advantages of increasing the baseline are tantalizing.  Increased detection sensitivity could allow for science reach beyond LISA's targets, potentially even giving access to signals of cosmological origin such as the predicted primordial gravitation waves generated by inflation \cite{Dimopoulos2008}.  In addition to substantially enhanced signal strength, the size of many background noise sources are suppressed.  Generally, a local acceleration noise source $\delta a$ results in an effective strain response $\propto \delta a / L$, so longer baselines can reduce the technical requirements need to control a wide class of backgrounds.  In addition, at the same target GW signal strength, increasing the baseline can reduce the need to use large momentum transfer (LMT) and other phase enhancement techniques \cite{Chiow2011, Graham2013a}, simplifying the interferometer operation.

\begin{figure*}
\centering
\includegraphics[width=\textwidth]{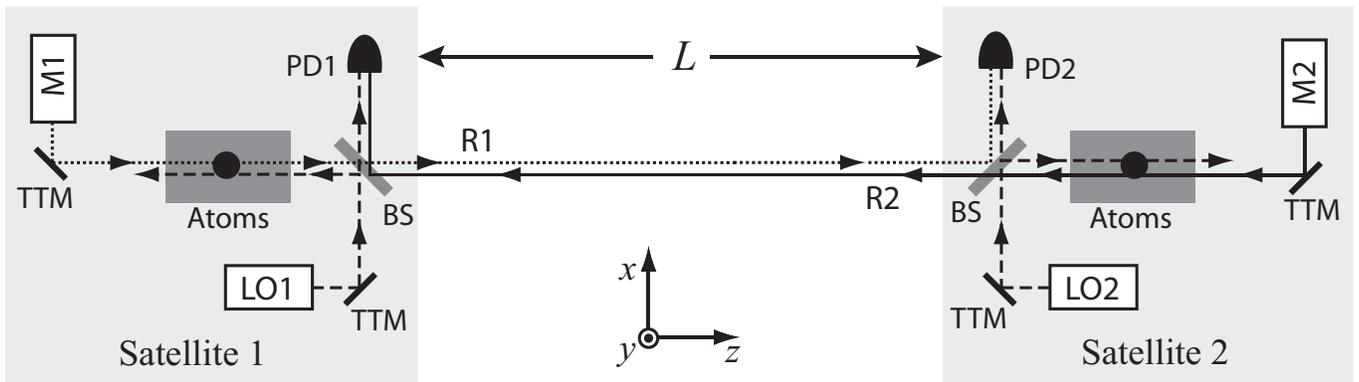}
\caption{ \label{Fig:apparatus} Schematic of the proposed design. M1 and M2 are the master lasers, with beams depicted as dotted and solid lines, respectively. The reference beams propagating between the satellites are denoted R1 (dotted) and R2 (solid).  LO1 and LO2 are local oscillator lasers (dashed beam lines) that are phase locked to the incoming reference laser beams (R2 and R1, respectively). PD1 (PD2) is a photodetector used to measure the heterodyne beatnote between the incoming reference beam R2 (R1) and the local oscillator laser LO1 (LO2) in order to provide feedback for the laser link. BS is a (non-polarizing) beamsplitter where the heterodyne beatnote is formed. Tip-tilt mirrors (TTM) allow for fine control of the pointing direction of each laser. All adjacent parallel beams are nominally overlapped, but for clarity they are shown here with a small offset.}
\end{figure*}

Here we propose a new concept for an atom interferometric GW detector that can support substantially longer baselines without requiring proportionally larger telescopes or increased laser power.  The idea is analogous to the traditional laser links used to connect the test masses in the LISA concept \cite{Bender1998,Gerardi2014}.  In our proposal, intense local lasers are used to operate the atom interferometers at each end of the baseline.  To connect these otherwise independent local lasers, reference lasers beams are transmitted between the two spacecraft, and the local lasers are kept phase locked to the incoming wavefronts of these reference lasers. In this scheme, the reference beams do not need to be collimated, since the phase locks can be done using much less intensity than is required to drive the atomic transitions.  This allows the baseline to be extended to LISA-like lengths with only a modest telescope size and reference beam power.  Critically, since the phase-locked local laser tracks the noise of the incoming reference laser, this arrangement maintains the essential common-mode laser phase noise cancellation between the two interferometers that allows for single baseline operation.  The current proposal effectively decouples the phase noise rejection requirement from the intensity demands, allowing the flexibility to independently optimize the baseline and atomic transition rate.

A schematic of the proposed design is shown in Fig. \ref{Fig:apparatus}.  Each satellite contains an atom interferometer that is implemented using laser pulses traveling along both the positive and negative $z$ direction.  Both satellites contain their own master laser (M1 and M2) that has enough intensity to drive transitions in the local atom interferometer.  After interacting with the local atom cloud, each master laser beam exits the satellite through a beamsplitter and then propagates across the baseline towards the opposite satellite.  We refer to the beams propagating between the satellites as reference beams: R1 and R2 are the reference beams originating from satellite 1 and 2, respectively.

The reference beams are not assumed to be collimated when they reach the opposite satellite, so for very long baselines the received reference beam intensity is expected to be too low to directly drive an atomic transition.  To address this, local oscillator lasers (LO1 and LO2) are phase locked to the incoming reference beams, and these lasers have sufficient intensity to drive transitions in the local atom interferometers.  The phase lock for laser LO1 is implemented by detecting the heterodyne beatnote formed by the incoming reference beam R2 with laser LO1 on the beam splitter BS in satellite 1 (and analogously for LO2 in satellite 2).  In addition to a photodetector for measuring the phase difference between the two beams, a quadrant detector (or camera) may be used to characterize the spatial interference pattern.  This allows the pointing direction and spatial mode of the two lasers to be well matched using appropriate feedback, as discussed later.

An essential consideration is the required noise performance of the phase lock between the reference beam and the local oscillator in each satellite. Any noise added by the phase lock is not common between the interferometers and so must be sufficiently small.  One source of noise could arise from motion of the beam splitter (BS).  The beam splitter is assumed to be rigidly connected to the satellite bus, so any platform vibration noise will affect the beam splitter as well.  Recall that in the original atom GW design \cite{Dimopoulos2008,Graham2013a}, light propagates across the baseline between the two atom ensembles without encountering any intervening optics, decoupling the differential atom signal from satellite platform accelerations.  In contrast, the reference beams in the current scheme interact with the beam splitters before reaching the atoms, conceivably adding noise.  However, it turns out that the proposed scheme is insensitive to phase noise introduced by vibration of the beam splitters.

To see this, assume that the incoming reference laser has phase $\phi_\text{R}$ at the nominal position $\v{r}=0$ of the beam splitter. Due to vibration of the satellite, the beam splitter may be displaced by some amount $\v{\Delta r}$, so upon reflection from the beam splitter the reference beam will instead have phase $\phi_\text{R}' = \phi_\text{R} + \v{k}\cdot\v{\Delta r}$, where $\v{k}$ is the wavevector of the incoming reference beam. On the other hand, the LO beam that transmits through the beam splitter is not impacted by the displacement, so the heterodyne signal between the lasers encodes the vibration noise $\v{\Delta r}$.  When the phase lock is engaged, the phase of the LO laser $\phi_\text{LO}$ at $\v{r}=0$ is locked to the phase of the reflected reference beam, resulting in $\phi_\text{LO} = \phi_\text{R} + \v{k}\cdot\v{\Delta r}$. Finally, consider the phase $\phi_\text{LO}'$ of the LO beam that reflects off the beam splitter and that is subsequently incident on the atoms.  Since the LO beam reflects off the opposite side of the beam splitter compared to the reference beam, the reflected LO beam at $\v{r}=0$ has phase $\phi_\text{LO}' = \phi_\text{LO} - \v{k}\cdot\v{\Delta r}$.  This implies that $\phi_\text{LO}'=\phi_\text{R}$ as desired, so the vibration noise does not affect the light reaching the atoms.

The phase lock is ultimately limited by photon shot noise of the received reference beam light. This is a new constraint that has not been present in past designs based on collimated beams.  The optical phase noise power spectral density (PSD) of the shot noise is approximately $S_\text{sh}=h \nu /P_r$, where $P_r$ is the power of the received reference beam and $\nu=c/\lambda$ is the light frequency.  Assuming a Gaussian beam with power $P_t$ and radial waist $w_t$ at the transmitting satellite, the received power collected by a telescope with diameter $d$ is approximately $P_r\approx\tfrac{1}{2}P_t(d/w_r)^2$ for $d\ll w_r$, where $w_r\approx L\lambda / \pi w_t$ is the reference beam waist after propagating a distance $L\gg z_\text{R}$ to the receiving satellite location.

To avoid limiting the detector strain resolution, the photon shot noise contribution to the atom interferometer phase must be less than the contribution from atom shot noise.  The RMS phase response of the atom interferometer to a laser pulse with optical phase noise PSD $S_\phi(\omega)$ is $\delta\phi_\text{rms}^2=\int \! S_\phi(\omega) |H(\omega)|^2 d\omega$, where $|H(\omega)|$ is the atom interferometer optical phase noise susceptibility \cite{Cheinet2008}. For a single pulse, the susceptibility is a low pass filter with $|H(\omega)|\approx 1$ out to roughly the Rabi frequency $\omega=\Omega$ \cite{Cheinet2008, Dimopoulos2008}.  Optical shot noise has a white power spectrum $S_\phi(\omega)=S_\text{sh}$, so the RMS phase noise for an interferometer with $n_p$ (assumed uncorrelated) pulses is approximately $\delta\phi_\text{rms}^2\approx n_p S_\text{sh}\Omega$. The Rabi frequency $\Omega=\sqrt{I_t / 2 I_\text{sat}}$ is set by the intensity $I_t = \tfrac{P_t}{\pi w_t^2/2}$ at the transmitting satellite and assumes a two-level transition with saturation intensity $I_\text{sat} = 2\pi^2\hbar c \, \Gamma/3\lambda^3$ and natural linewidth $\Gamma$.  Assuming an atom interferometer repetition rate of $f_R$, the associated noise PSD is $\overline{\delta\phi}_\gamma^2=\delta\phi_\text{rms}^2/f_R$. The final noise amplitude spectral density is then
\be
\overline{\delta\phi}_\gamma = \sqrt[4]{\frac{1536\,\hbar c }{\pi^5}} \, \frac{n_p^{1/2} \Gamma^{1/4} \lambda^{5/4} L}{f_R^{1/2} P_t^{1/4} d^{5/2}}
\label{Eq:PhotoShotNoise}
\ee
where the telescope diameter is taken to be $d=2 w_t$.  By comparison, the phase noise PSD of atom shot noise is $\overline{\delta\phi}_a^2=1/N_a$, where $N_a$ is the mean number of detected atoms per unit time that participate in the atom interferometer signal.  In designing the laser link phase lock we require that $\overline{\delta\phi}_\gamma\leq\overline{\delta\phi}_a$.  Assuming the Sr clock transition, the telescope diameter is constrained to be
\be
d=28~\text{cm}\,\Big(\!\tfrac{L}{2\cdot10^9~\!\text{m}}\!\Big)^{\!\frac{2}{5}}  \Big(\!\tfrac{1~\!\text{W}}{P_t}\!\Big)^{\!\frac{1}{10}}  \Big(\!\tfrac{0.2~\!\text{Hz}/7}{f_R/n_p}\!\Big)^{\!\frac{1}{5}}  \Big(\!\tfrac{\overline{\delta\phi}_a}{10^{-3}/\sqrt{\text{Hz}}}\!\Big)^{\!\frac{2}{5}}
\ee
where $n_p=7$ corresponds to a $2\hbar k$ interferometer \cite{Graham2013a}.

\begin{figure}
\centering
\includegraphics[width=\columnwidth]{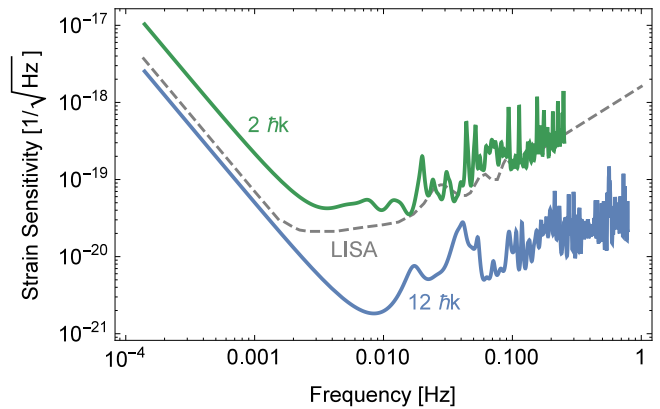}
\caption{ \label{Fig:Sensitivity} Strain sensitivity for a $2\hbar k$ Sr interferometer with baseline $L=2\times 10^9~\text{m}$ (green) as well as a $12\hbar k$ interferometer with baseline $L=6\times 10^8~\text{m}$ (blue). The strain responses have been averaged over gravitational wave propagation direction and polarization. The $2\hbar k$ curve represents an average of three alternating interferometer interrogation times: $T=160~\text{s}, 100~\text{s},40~\text{s}$. The $12\hbar k$ curve is an average of $T=75~\text{s}, 69~\text{s},59~\text{s},53~\text{s}$, limited at the low end by the light travel time. The interrogation time average regularizes the detector response by suppressing well known notches in single interferometer transfer functions \cite{Dimopoulos2008}. The LISA strain curve is shown for reference.}
\end{figure}

Figure~\ref{Fig:Sensitivity} shows the strain sensitivity curves for two long-baseline designs using Sr atoms. The more conservative design (green) uses an $L=2\times 10^9~\text{m}$ baseline and the photon shot-noise limited laser link assumes $1~\text{W}$ laser power, a $d=30~\text{cm}$ diameter telescope, as well as $\chi=50$ concurrent interferometers to support a $f_R=\chi/2T\approx 0.2~\text{Hz}$ sampling rate.  The long baseline allows for high sensitivity even though the design assumes conservative $2\hbar k$ atom optics and atom shot-noise of $\overline{\delta\phi}_a = 10^{-3}~\text{rad}/\sqrt{\text{Hz}}$.  A long interrogation time of $T=160~\text{s}$ is used to support low frequency sensitivity, but despite this long drift time the maximum wavepacket separation is bounded to $<2~\text{m}$, minimizing satellite size.  The atom source design assumes ensembles of $7\times 10^6$ atoms with a $20~\text{pK}$ longitudinal temperature, allowing for a $\Omega/2\pi=60~\text{Hz}$ Rabi frequency.  Such design criteria are readily met using existing technology \cite{Kovachy2014}.

LMT techniques allow for enhanced sensitivity, as shown by the second strain sensitivity curve (blue) in Fig.~\ref{Fig:Sensitivity}. This design is based on a $12\hbar k$ interferometer sequence with an $L=6\times 10^8~\text{m}$ baseline and improved phase noise $\overline{\delta\phi}_a = 10^{-4}~\text{rad}/\sqrt{\text{Hz}}$. Photon shot noise requirements are met using $1~\text{W}$ laser power and a $d=50~\text{cm}$ diameter telescope, giving Rabi frequency $\Omega/2\pi=40~\text{Hz}$.  The design has a sampling rate of $f_R=\chi/2T\approx 1~\text{Hz}$ and assumes $\chi=120$ concurrent interferometers \footnote{Concurrent interferometers can be implemented using different initial velocities so that each interferometer is addressable with a unique Doppler shift. Strategies for efficient allocation of Doppler bandwidth for LMT interferometers are discussed in \cite{multiplex}.}.  The increased phase sensitivity of this design allows for improved low frequency response even using a smaller interrogation time.  Using $T=75~\text{s}$, the maximum wavepacket separation is $<4~\text{m}$.

Another source of noise is timing delay and jitter in the pulses emitted by the phase-locked local oscillator laser. Referring to Fig.~\ref{Fig:apparatus}, consider a pulse emitted from M1 at time $t$ that arrives at the satellite 2 beam splitter at time $t_a$.  If the pulse emitted from LO2 is delayed by some time $t_d$ after the arrival of the reference pulse, the laser phase $\phi_{LO}(t_a+t_d)$ imprinted on interferometer 2 will be different from the phase $\phi_{M}(t)$ written onto interferometer 1 by the amount $\phi_{LO}(t_a+t_d)-\phi_{M}(t) = \omega t_d + \delta\phi$, where $\omega$ is the instantaneous frequency of laser LO2 at time $t_a$ and $\delta\phi \equiv \phi_{LO}(t_a) - \phi_{M}(t)$ quantifies any imperfection in the phase lock between R1 and LO2. In addition, the delayed transition in interferometer 2 implies that interferometer 1 spends more time in the excited state by comparison, leading to an extra differential phase $\omega_A t_d$ for excited state energy $\hbar \omega_A$ \footnote{The description here is of a transition from the ground to excited state. For transitions from excited to ground, the signs of all the phase shifts are flipped.}. The total gradiometer phase due to the delayed pulse is then $\Delta\phi_\text{delay}=(\omega-\omega_A)t_d + \delta\phi$. Assuming a perfect phase lock ($\delta\phi=0$), noise can arise from either timing jitter $\overline{\delta t}_d$ or frequency noise $\overline{\delta\omega}$, giving a noise PSD of $\overline{\delta\phi}_\text{delay}^2=(N \, t_d \,\overline{\delta\omega})^2 + (N \, \Delta \,\overline{\delta t}_d)^2$, where $\Delta = \omega-\omega_A$ is the pulse detuning and $N$ is any LMT phase enhancement factor. Keeping each term below $\overline{\delta\phi}_a$ requires noise amplitude spectral densities of
\begin{align}
\overline{\delta\omega} =&\, 2\pi\!\times\! 80~\!\tfrac{\text{Hz}}{\sqrt{\text{Hz}}}\,\, \Big(\!\tfrac{2}{N}\!\Big)\!\left(\!\tfrac{1~\text{$\mu$s}}{t_d}\!\right)\!\left(\!\tfrac{\overline{\delta\phi}_a}{10^{-3}~\text{rad}/\sqrt{\text{Hz}}}\!\right)\\
\overline{\delta t}_d =&\, 1.3~\!\tfrac{\text{$\mu$s}}{\sqrt{\text{Hz}}}\,\, \Big(\!\tfrac{2}{N}\!\Big)\!\left(\tfrac{60~\text{Hz}}{\Delta/2\pi}\right)\!\left(\!\tfrac{\overline{\delta\phi}_a}{10^{-3}~\text{rad}/\sqrt{\text{Hz}}}\!\right)
\end{align}
at frequencies in the GW detection band. In particular, this shows that the long-time frequency stability requirements of the LO laser can be reduced by ensuring that the pulses from the LO are well synchronized with the incoming reference pulses, keeping $t_d$ small.  In practice, $t_d\sim 10~\text{ns}$ with RMS noise $\delta t_d\sim 1~\text{ns}$ appears straightforward, suggesting that LO pulse timing constraints are manageable.

Satellite and laser beam pointing jitter can also introduce noise.  Consider a beam propagating approximately along the $z$ axis in Fig.~\ref{Fig:apparatus} from satellite 1 towards satellite 2 that is tilted by a small angle $\theta_y$ about the $y$-axis.  Near satellite 1, the phase of the Gaussian beam varies with position as $\Phi_1(x,z)\approx k z + k \theta_y x$, where the center of rotation of the beam is taken to be $x=0$, $z=0$. By comparison, the phase near satellite 2 is $\Phi_2(x,z)\approx k (z + L) + k \theta_y (z_R/L)^2 x$ for baseline length $L$ much longer than the Rayleigh range $z_R$.  Here a long baseline is advantageous since when $z_R\ll L$ the beam arriving at satellite 2 is approximately a spherical wave, so the dependance of the phase on angle is greatly suppressed.  In this limit, the pointing jitter constraint is set by the $\Phi_1$ coupling and has noise amplitude $\overline{\delta \phi}_{\theta}= 4 k N \Delta x \, \overline{\delta\theta}$, where $\overline{\delta\theta}^2$ is the angle noise PSD and $\Delta x$ is the transverse position offset of the atom relative to the baseline \cite{calcAssumptions, Hogan2011}. The pointing requirement is then
\be\overline{\delta\theta} = 10~\!\tfrac{\text{nrad}}{\sqrt{\text{Hz}}} \, \Big(\!\tfrac{2}{N}\!\Big)\!\Big(\!\tfrac{1~\text{mm}}{\Delta x}\!\Big)\!\left(\!\tfrac{\overline{\delta\phi}_a}{10^{-3}~\text{rad}/\sqrt{\text{Hz}}}\!\right).\label{Eq:pointingReq}\ee

To avoid introducing additional pointing noise, the LO laser beam incident on the atoms must point in the same direction as the incoming reference laser pulse. This can be facilitated by monitoring the relative angle between the two beams at the beam splitter. In addition to measuring the beat note for the phase lock, a position sensitive detector such as as quadrant photodiode or a CCD camera could be used to record the spatial interference pattern between the reference and LO beams. Feedback applied to a tip-tilt mirror (show as TTM in Fig.~\ref{Fig:apparatus} before the BS) can then be used to control the angle of the LO laser. Similarly, the angle of the master laser itself can be controlled by comparing it to LO laser direction and using another tip-tilt mirror. The interference signal between the LO and the master can be generated using a Michelson interferometer geometry, inserting an additional beam splitter at any point along the path where the two beams are counter-propagating.  In this configuration, the pointing stability of all the beams is tied to the stability of the incoming (nearly) spherical wavefronts of the reference beam.

The performance of the angle control loop is ultimately limited by the shot noise of the received reference wavefront. To estimate this, note that the power difference $\Delta P$ between the two sides of a quadrant detector due to the spatial interference pattern caused by a small angle $\Delta\theta$ between the LO and the reference beam is $\Delta P \approx 4\sqrt{2\pi}\sqrt{P_{LO}P_{r}} \Delta\theta\, w_t/\lambda$ for a telescope diameter $d=2 w_t$ and received powers $P_{LO}$ and $P_r$ from the two beams \cite{angleMeasurement}.  The noise in $\Delta P$ is dominated by the strong LO beam, giving a noise PSD of $\overline{\delta(\Delta P)}^2\approx \overline{\delta P}_{LO}^2 = h \nu P_{LO}$ assuming shot noise for the LO optical power noise. The shot noise limit for a measurement of $\Delta\theta$ is then given by amplitude spectral density
\be\overline{\delta(\Delta\theta)} = \tfrac{\sqrt{S_\text{sh}}}{4\sqrt{2\pi}}\tfrac{\lambda}{w_t} \approx 1~\!\tfrac{\text{nrad}}{\sqrt{\text{Hz}}}\,\Big(\!\tfrac{10~\text{cm}}{w_t}\!\Big)\!\left(\!\tfrac{\overline{\delta\phi}_a}{10^{-3}~\text{rad}/\sqrt{\text{Hz}}}\!\right)\ee
where $S_\text{sh}=h\nu/P_r$ is the phase noise PSD of the reference beam and we assume a design with $S_\text{sh}=\overline{\delta\phi}_a^2$ as before. Comparing this with the requirement in Eq.~\ref{Eq:pointingReq} suggests that the angle can be sufficiently well measured to control the LO pointing direction. This also suggests that overall satellite bus pointing requirements are modest ($\sim 10^{-6}~\text{rad}/\sqrt{\text{Hz}}$, limited by the dynamic range of the pointing servos), and substantially reduced from previous proposals.

Past designs have required large momentum transfer to reach design sensitivity.  The designs proposed here operate at lower momentum transfer and thus place much less stringent constraints on laser phase front stability.  For example, here we require a phase stability of $\lambda/30$ for the telescope. Following references \cite{Hogan2011, Dimopoulos2008, Bender2011, Dimopoulos2011}, laser wavefront aberrations $\delta\lambda/\lambda$ couple to satellite transverse position noise $\overline{\delta x}$, resulting in phase noise amplitude $\overline{\delta \phi}_\lambda = 16\pi^2 N (\delta\lambda/\lambda) \overline{\delta x} / \Lambda$, where $\Lambda$ is the aberration wavelength \cite{calcAssumptions}. The estimated wavefront requirement for the interferometer beam is then \cite{aberrationNoise}
\be\delta\lambda=\tfrac{\lambda}{30} \, \Big(\!\tfrac{2}{N}\!\Big)\!\Big(\!\tfrac{\Lambda}{1~\text{cm}}\!\Big)\!\left(\!\tfrac{\overline{\delta\phi}_a}{10^{-3}~\text{rad}/\sqrt{\text{Hz}}}\!\right)\left(\!\tfrac{\overline{\delta x}}{1~\text{$\mu$m}/\sqrt{\text{Hz}}}\!\right)\ee
where we have assumed satellite transverse position jitter of $\overline{\delta x} = 1~\text{$\mu$m}/\sqrt{\text{Hz}}$.  This suggests modest satellite bus jitter requirements.

In addition to pointing errors, measuring the spatial interference pattern can also provide information about the wavefront of the LO laser. Since most abberations have diffracted out of the beam after propagating across the long baseline, the reference beam is a pristine optical wavefront reference. It might be possible to mitigate wavefront requirements further \cite{Hogan2011} by combining this wavefront measurement with appropriate feedback on the LO mode cleaning optics.

Instrument constraints imposed by backgrounds that have their origin in spurious forces or phase shifts  (due to, for example, magnetic field gradients, blackbody shifts, AC Stark shifts or gravitational gradients) are significantly eased due to the longer baseline.  Since previous shorter baseline designs could meet these requirements \footnote{See, for example, the analysis in Refs.~\cite{Hogan2011,Graham2013a}}, in the longer baseline designs proposed here these backgrounds can be brought to levels where they do not impact instrument performance.

Several variations of the basic laser link concept are possible.  For example, by employing a second laser link on each satellite, a dedicated reference laser can be used to generate the reference beams instead of relying on the transmitted master laser light as shown in Fig.~\ref{Fig:apparatus}. In this modified setup, the heterodyne signal between the master laser and the new reference laser is formed on the existing beam splitter (BS) such that the reflected reference laser beam is directed towards the opposite satellite.  Additional beam splitters, samplers, and polarization optics may be used to overlap the new beam paths with the existing LO lock paths.

Generalizing further, if the reference beam is a separate laser then in principle its wavelength can be different from that of the atomic transition. An optical frequency comb would then be used to implement the heterodyne lock, spanning the frequency difference between the reference laser and the lasers responsible for interrogating the atoms. Changing the reference wavelength could lead to lower optical shot noise ($\sim\lambda^{5/4}$ in Eq.~\ref{Eq:PhotoShotNoise}) or could exploit existing laser technology at particular wavelengths.

\bibliographystyle{apsrev4-1}
\bibliography{heterodyneGW,heterodyneGWExtraCites}

\begin{thebibliography}{20}%
\makeatletter
\providecommand \@ifxundefined [1]{%
 \@ifx{#1\undefined}
}%
\providecommand \@ifnum [1]{%
 \ifnum #1\expandafter \@firstoftwo
 \else \expandafter \@secondoftwo
 \fi
}%
\providecommand \@ifx [1]{%
 \ifx #1\expandafter \@firstoftwo
 \else \expandafter \@secondoftwo
 \fi
}%
\providecommand \natexlab [1]{#1}%
\providecommand \enquote  [1]{``#1''}%
\providecommand \bibnamefont  [1]{#1}%
\providecommand \bibfnamefont [1]{#1}%
\providecommand \citenamefont [1]{#1}%
\providecommand \href@noop [0]{\@secondoftwo}%
\providecommand \href [0]{\begingroup \@sanitize@url \@href}%
\providecommand \@href[1]{\@@startlink{#1}\@@href}%
\providecommand \@@href[1]{\endgroup#1\@@endlink}%
\providecommand \@sanitize@url [0]{\catcode `\\12\catcode `\$12\catcode
  `\&12\catcode `\#12\catcode `\^12\catcode `\_12\catcode `\%12\relax}%
\providecommand \@@startlink[1]{}%
\providecommand \@@endlink[0]{}%
\providecommand \url  [0]{\begingroup\@sanitize@url \@url }%
\providecommand \@url [1]{\endgroup\@href {#1}{\urlprefix }}%
\providecommand \urlprefix  [0]{URL }%
\providecommand \Eprint [0]{\href }%
\providecommand \doibase [0]{http://dx.doi.org/}%
\providecommand \selectlanguage [0]{\@gobble}%
\providecommand \bibinfo  [0]{\@secondoftwo}%
\providecommand \bibfield  [0]{\@secondoftwo}%
\providecommand \translation [1]{[#1]}%
\providecommand \BibitemOpen [0]{}%
\providecommand \bibitemStop [0]{}%
\providecommand \bibitemNoStop [0]{.\EOS\space}%
\providecommand \EOS [0]{\spacefactor3000\relax}%
\providecommand \BibitemShut  [1]{\csname bibitem#1\endcsname}%
\let\auto@bib@innerbib\@empty
\bibitem [{\citenamefont {Dimopoulos}\ \emph {et~al.}(2008)\citenamefont
  {Dimopoulos}, \citenamefont {Graham}, \citenamefont {Hogan}, \citenamefont
  {Kasevich},\ and\ \citenamefont {Rajendran}}]{Dimopoulos2008}%
  \BibitemOpen
  \bibfield  {author} {\bibinfo {author} {\bibfnamefont {S.}~\bibnamefont
  {Dimopoulos}}, \bibinfo {author} {\bibfnamefont {P.}~\bibnamefont {Graham}},
  \bibinfo {author} {\bibfnamefont {J.}~\bibnamefont {Hogan}}, \bibinfo
  {author} {\bibfnamefont {M.}~\bibnamefont {Kasevich}}, \ and\ \bibinfo
  {author} {\bibfnamefont {S.}~\bibnamefont {Rajendran}},\ }\href {\doibase
  10.1103/PhysRevD.78.122002} {\bibfield  {journal} {\bibinfo  {journal}
  {Physical Review D}\ }\textbf {\bibinfo {volume} {78}},\ \bibinfo {pages}
  {122002} (\bibinfo {year} {2008})}\BibitemShut {NoStop}%
\bibitem [{\citenamefont {Hogan}\ \emph
  {et~al.}(2011{\natexlab{a}})\citenamefont {Hogan}, \citenamefont {Johnson},
  \citenamefont {Dickerson}, \citenamefont {Kovachy}, \citenamefont
  {Sugarbaker}, \citenamefont {Chiow}, \citenamefont {Graham}, \citenamefont
  {Kasevich}, \citenamefont {Saif}, \citenamefont {Rajendran}, \citenamefont
  {Bouyer}, \citenamefont {Seery}, \citenamefont {Feinberg},\ and\
  \citenamefont {Keski-Kuha}}]{Hogan2011}%
  \BibitemOpen
  \bibfield  {author} {\bibinfo {author} {\bibfnamefont {J.~M.}\ \bibnamefont
  {Hogan}}, \bibinfo {author} {\bibfnamefont {D.~M.~S.}\ \bibnamefont
  {Johnson}}, \bibinfo {author} {\bibfnamefont {S.}~\bibnamefont {Dickerson}},
  \bibinfo {author} {\bibfnamefont {T.}~\bibnamefont {Kovachy}}, \bibinfo
  {author} {\bibfnamefont {A.}~\bibnamefont {Sugarbaker}}, \bibinfo {author}
  {\bibfnamefont {S.-w.}\ \bibnamefont {Chiow}}, \bibinfo {author}
  {\bibfnamefont {P.~W.}\ \bibnamefont {Graham}}, \bibinfo {author}
  {\bibfnamefont {M.~A.}\ \bibnamefont {Kasevich}}, \bibinfo {author}
  {\bibfnamefont {B.}~\bibnamefont {Saif}}, \bibinfo {author} {\bibfnamefont
  {S.}~\bibnamefont {Rajendran}}, \bibinfo {author} {\bibfnamefont
  {P.}~\bibnamefont {Bouyer}}, \bibinfo {author} {\bibfnamefont {B.~D.}\
  \bibnamefont {Seery}}, \bibinfo {author} {\bibfnamefont {L.}~\bibnamefont
  {Feinberg}}, \ and\ \bibinfo {author} {\bibfnamefont {R.}~\bibnamefont
  {Keski-Kuha}},\ }\href {\doibase 10.1007/s10714-011-1182-x} {\bibfield
  {journal} {\bibinfo  {journal} {General Relativity and Gravitation}\ }\textbf
  {\bibinfo {volume} {43}},\ \bibinfo {pages} {1953} (\bibinfo {year}
  {2011}{\natexlab{a}})}\BibitemShut {NoStop}%
\bibitem [{Note1()}]{Note1}%
  \BibitemOpen
  \bibinfo {note} {The properties of the atomic proof-masses are regular (all
  atoms are identical) and knowable (well defined magnetic susceptibility).
  They are immune from charging events.}\BibitemShut {Stop}%
\bibitem [{\citenamefont {Graham}\ \emph {et~al.}(2013)\citenamefont {Graham},
  \citenamefont {Hogan}, \citenamefont {Kasevich},\ and\ \citenamefont
  {Rajendran}}]{Graham2013a}%
  \BibitemOpen
  \bibfield  {author} {\bibinfo {author} {\bibfnamefont {P.~W.}\ \bibnamefont
  {Graham}}, \bibinfo {author} {\bibfnamefont {J.~M.}\ \bibnamefont {Hogan}},
  \bibinfo {author} {\bibfnamefont {M.~A.}\ \bibnamefont {Kasevich}}, \ and\
  \bibinfo {author} {\bibfnamefont {S.}~\bibnamefont {Rajendran}},\ }\href
  {\doibase 10.1103/PhysRevLett.110.171102} {\bibfield  {journal} {\bibinfo
  {journal} {Physical Review Letters}\ }\textbf {\bibinfo {volume} {110}},\
  \bibinfo {pages} {171102} (\bibinfo {year} {2013})}\BibitemShut {NoStop}%
\bibitem [{\citenamefont {Kasevich}\ and\ \citenamefont
  {Chu}(1991)}]{Kasevich1991}%
  \BibitemOpen
  \bibfield  {author} {\bibinfo {author} {\bibfnamefont {M.}~\bibnamefont
  {Kasevich}}\ and\ \bibinfo {author} {\bibfnamefont {S.}~\bibnamefont {Chu}},\
  }\href {\doibase 10.1103/PhysRevLett.67.181} {\bibfield  {journal} {\bibinfo
  {journal} {Physical Review Letters}\ }\textbf {\bibinfo {volume} {67}},\
  \bibinfo {pages} {181} (\bibinfo {year} {1991})}\BibitemShut {NoStop}%
\bibitem [{\citenamefont {Chiow}\ \emph {et~al.}(2011)\citenamefont {Chiow},
  \citenamefont {Kovachy}, \citenamefont {Chien},\ and\ \citenamefont
  {Kasevich}}]{Chiow2011}%
  \BibitemOpen
  \bibfield  {author} {\bibinfo {author} {\bibfnamefont {S.-w.}\ \bibnamefont
  {Chiow}}, \bibinfo {author} {\bibfnamefont {T.}~\bibnamefont {Kovachy}},
  \bibinfo {author} {\bibfnamefont {H.-C.}\ \bibnamefont {Chien}}, \ and\
  \bibinfo {author} {\bibfnamefont {M.~A.}\ \bibnamefont {Kasevich}},\ }\href
  {\doibase 10.1103/PhysRevLett.107.130403} {\bibfield  {journal} {\bibinfo
  {journal} {Physical Review Letters}\ }\textbf {\bibinfo {volume} {107}},\
  \bibinfo {pages} {130403} (\bibinfo {year} {2011})}\BibitemShut {NoStop}%
\bibitem [{\citenamefont {{LISA Study Team}}(1998)}]{Bender1998}%
  \BibitemOpen
  \bibfield  {author} {\bibinfo {author} {\bibnamefont {{LISA Study Team}}},\
  }\href {http://lisa.nasa.gov/Documentation/ppa2.08.pdf} {\enquote {\bibinfo
  {title} {{LISA Pre-Phase A Report}},}\ } (\bibinfo {year} {1998})\BibitemShut
  {NoStop}%
\bibitem [{\citenamefont {Gerardi}\ \emph {et~al.}(2014)\citenamefont
  {Gerardi}, \citenamefont {Allen}, \citenamefont {Conklin}, \citenamefont
  {Sun}, \citenamefont {DeBra}, \citenamefont {Buchman}, \citenamefont {Gath},
  \citenamefont {Fichter}, \citenamefont {Byer},\ and\ \citenamefont
  {Johann}}]{Gerardi2014}%
  \BibitemOpen
  \bibfield  {author} {\bibinfo {author} {\bibfnamefont {D.}~\bibnamefont
  {Gerardi}}, \bibinfo {author} {\bibfnamefont {G.}~\bibnamefont {Allen}},
  \bibinfo {author} {\bibfnamefont {J.~W.}\ \bibnamefont {Conklin}}, \bibinfo
  {author} {\bibfnamefont {K.-X.}\ \bibnamefont {Sun}}, \bibinfo {author}
  {\bibfnamefont {D.}~\bibnamefont {DeBra}}, \bibinfo {author} {\bibfnamefont
  {S.}~\bibnamefont {Buchman}}, \bibinfo {author} {\bibfnamefont
  {P.}~\bibnamefont {Gath}}, \bibinfo {author} {\bibfnamefont {W.}~\bibnamefont
  {Fichter}}, \bibinfo {author} {\bibfnamefont {R.~L.}\ \bibnamefont {Byer}}, \
  and\ \bibinfo {author} {\bibfnamefont {U.}~\bibnamefont {Johann}},\ }\href
  {\doibase 10.1063/1.4862199} {\bibfield  {journal} {\bibinfo  {journal} {The
  Review of Scientific Instruments}\ }\textbf {\bibinfo {volume} {85}},\
  \bibinfo {pages} {011301} (\bibinfo {year} {2014})}\BibitemShut {NoStop}%
\bibitem [{\citenamefont {Cheinet}\ \emph {et~al.}(2008)\citenamefont
  {Cheinet}, \citenamefont {Canuel}, \citenamefont {{Pereira Dos Santos}},
  \citenamefont {Gauguet}, \citenamefont {Yver-Leduc},\ and\ \citenamefont
  {Landragin}}]{Cheinet2008}%
  \BibitemOpen
  \bibfield  {author} {\bibinfo {author} {\bibfnamefont {P.}~\bibnamefont
  {Cheinet}}, \bibinfo {author} {\bibfnamefont {B.}~\bibnamefont {Canuel}},
  \bibinfo {author} {\bibfnamefont {F.}~\bibnamefont {{Pereira Dos Santos}}},
  \bibinfo {author} {\bibfnamefont {A.}~\bibnamefont {Gauguet}}, \bibinfo
  {author} {\bibfnamefont {F.}~\bibnamefont {Yver-Leduc}}, \ and\ \bibinfo
  {author} {\bibfnamefont {A.}~\bibnamefont {Landragin}},\ }\href {\doibase
  10.1109/TIM.2007.915148} {\bibfield  {journal} {\bibinfo  {journal} {IEEE
  Transactions on Instrumentation and Measurement}\ }\textbf {\bibinfo {volume}
  {57}},\ \bibinfo {pages} {1141} (\bibinfo {year} {2008})}\BibitemShut
  {NoStop}%
\bibitem [{\citenamefont {Kovachy}\ \emph {et~al.}(2014)\citenamefont
  {Kovachy}, \citenamefont {Hogan}, \citenamefont {Sugarbaker}, \citenamefont
  {Dickerson}, \citenamefont {Donnelly}, \citenamefont {Overstreet},\ and\
  \citenamefont {Kasevich}}]{Kovachy2014}%
  \BibitemOpen
  \bibfield  {author} {\bibinfo {author} {\bibfnamefont {T.}~\bibnamefont
  {Kovachy}}, \bibinfo {author} {\bibfnamefont {J.~M.}\ \bibnamefont {Hogan}},
  \bibinfo {author} {\bibfnamefont {A.}~\bibnamefont {Sugarbaker}}, \bibinfo
  {author} {\bibfnamefont {S.~M.}\ \bibnamefont {Dickerson}}, \bibinfo {author}
  {\bibfnamefont {C.~A.}\ \bibnamefont {Donnelly}}, \bibinfo {author}
  {\bibfnamefont {C.}~\bibnamefont {Overstreet}}, \ and\ \bibinfo {author}
  {\bibfnamefont {M.~A.}\ \bibnamefont {Kasevich}},\ }\href
  {http://arxiv.org/abs/1407.6995} {\  (\bibinfo {year} {2014})},\ \Eprint
  {http://arxiv.org/abs/1407.6995} {arXiv:1407.6995} \BibitemShut {NoStop}%
\bibitem [{Note2()}]{Note2}%
  \BibitemOpen
  \bibinfo {note} {Concurrent interferometers can be implemented using
  different initial velocities so that each interferometer is addressable with
  a unique Doppler shift. Strategies for efficient allocation of Doppler
  bandwidth for LMT interferometers are discussed in \cite
  {multiplex}.}\BibitemShut {Stop}%
\bibitem [{Note3()}]{Note3}%
  \BibitemOpen
  \bibinfo {note} {The description here is of a transition from the ground to
  excited state. For transitions from excited to ground, the signs of all the
  phase shifts are flipped.}\BibitemShut {Stop}%
\bibitem [{cal()}]{calcAssumptions}%
  \BibitemOpen
  \href@noop {} {}\bibinfo {note} {Assumes a conventional $\pi/2-\pi-\pi/2$
  $N\hbar k$ LMT sequence and a perturbation frequency $\omega=\pi/T$ equal to
  the peak GW response frequency.}\BibitemShut {Stop}%
\bibitem [{ang()}]{angleMeasurement}%
  \BibitemOpen
  \href@noop {} {}\bibinfo {note} {See, for example,
  \cite{Hogan2011a}.}\BibitemShut {Stop}%
\bibitem [{\citenamefont {Bender}(2011)}]{Bender2011}%
  \BibitemOpen
  \bibfield  {author} {\bibinfo {author} {\bibfnamefont {P.~L.}\ \bibnamefont
  {Bender}},\ }\href {\doibase 10.1103/PhysRevD.84.028101} {\bibfield
  {journal} {\bibinfo  {journal} {Physical Review D}\ }\textbf {\bibinfo
  {volume} {84}},\ \bibinfo {pages} {028101} (\bibinfo {year}
  {2011})}\BibitemShut {NoStop}%
\bibitem [{\citenamefont {Dimopoulos}\ \emph {et~al.}(2011)\citenamefont
  {Dimopoulos}, \citenamefont {Graham}, \citenamefont {Hogan}, \citenamefont
  {Kasevich},\ and\ \citenamefont {Rajendran}}]{Dimopoulos2011}%
  \BibitemOpen
  \bibfield  {author} {\bibinfo {author} {\bibfnamefont {S.}~\bibnamefont
  {Dimopoulos}}, \bibinfo {author} {\bibfnamefont {P.~W.}\ \bibnamefont
  {Graham}}, \bibinfo {author} {\bibfnamefont {J.~M.}\ \bibnamefont {Hogan}},
  \bibinfo {author} {\bibfnamefont {M.~A.}\ \bibnamefont {Kasevich}}, \ and\
  \bibinfo {author} {\bibfnamefont {S.}~\bibnamefont {Rajendran}},\ }\href
  {\doibase 10.1103/PhysRevD.84.028102} {\bibfield  {journal} {\bibinfo
  {journal} {Physical Review D}\ }\textbf {\bibinfo {volume} {84}},\ \bibinfo
  {pages} {028102} (\bibinfo {year} {2011})}\BibitemShut {NoStop}%
\bibitem [{abe()}]{aberrationNoise}%
  \BibitemOpen
  \href@noop {} {}\bibinfo {note} {See Eq.~6 of Ref.~\cite{Hogan2011}. Here we
  neglect averaging over the cloud. Averaging would further relax requirements
  at short wavelengths.}\BibitemShut {Stop}%
\bibitem [{Note4()}]{Note4}%
  \BibitemOpen
  \bibinfo {note} {See, for example, the analysis in Refs.~\cite
  {Hogan2011,Graham2013a}}\BibitemShut {NoStop}%
\bibitem [{mul()}]{multiplex}%
  \BibitemOpen
  \href@noop {} {}\bibinfo {note} {J.M. Hogan, (to be published).}\BibitemShut
  {Stop}%
\bibitem [{\citenamefont {Hogan}\ \emph
  {et~al.}(2011{\natexlab{b}})\citenamefont {Hogan}, \citenamefont {Hammer},
  \citenamefont {Chiow}, \citenamefont {Dickerson}, \citenamefont {Johnson},
  \citenamefont {Kovachy}, \citenamefont {Sugarbaker},\ and\ \citenamefont
  {Kasevich}}]{Hogan2011a}%
  \BibitemOpen
  \bibfield  {author} {\bibinfo {author} {\bibfnamefont {J.~M.}\ \bibnamefont
  {Hogan}}, \bibinfo {author} {\bibfnamefont {J.}~\bibnamefont {Hammer}},
  \bibinfo {author} {\bibfnamefont {S.-W.}\ \bibnamefont {Chiow}}, \bibinfo
  {author} {\bibfnamefont {S.}~\bibnamefont {Dickerson}}, \bibinfo {author}
  {\bibfnamefont {D.~M.~S.}\ \bibnamefont {Johnson}}, \bibinfo {author}
  {\bibfnamefont {T.}~\bibnamefont {Kovachy}}, \bibinfo {author} {\bibfnamefont
  {A.}~\bibnamefont {Sugarbaker}}, \ and\ \bibinfo {author} {\bibfnamefont
  {M.~A.}\ \bibnamefont {Kasevich}},\ }\href
  {http://www.ncbi.nlm.nih.gov/pubmed/21540973} {\bibfield  {journal} {\bibinfo
   {journal} {Optics letters}\ }\textbf {\bibinfo {volume} {36}},\ \bibinfo
  {pages} {1698} (\bibinfo {year} {2011}{\natexlab{b}})}\BibitemShut {NoStop}%
\end{thebibliography}%

\end{document}